\documentclass[preprint]{elsart}
\usepackage[dvips]{graphicx}

\newcommand{\be}{\begin{equation}}
\newcommand{\ee}{\end{equation}}
\newcommand{\bea}{\vspace{0.25cm}\begin{eqnarray}}
\newcommand{\eea}{\end{eqnarray}}

\def\PLA{{Phys. Lett.}  A }

\def\PRA{{Phys. Rev.} A }

\begin{document}

\centerline{ \large \bf A conclusive experiment to throw more
light on "light"}

\vskip 1.2cm \centerline{ G. Brida, M. Genovese
\footnote{genovese@ien.it}, M. Gramegna} \vskip 0.5cm
\centerline{\it  Istituto Elettrotecnico Nazionale Galileo
Ferraris,} \centerline{ \it Strada delle Cacce 91, 10135 Torino,
Italy } \vskip 0.5cm \centerline{ E. Predazzi}\centerline{ \it
Dip. Fisica Teorica Univ. Torino and INFN, via P. Giuria 1, 10125
Torino, Italy }

\centerline{ \bf Abstract} \vskip 0.2cm

We describe a new realization of Ghose, Home, Agarwal experiment
on wave particle duality of light where some limitations of the
former experiment, realized by Mizobuchi and Ohtak\'e, are
overcome. Our results clearly indicate that wave-particle
complementarity must be understood between interference and
"whelcher weg" knowledge and not in a more general sense.


\vskip 1cm \vskip 1cm

Wave particle duality is one of the fundamental  aspects of
Quantum Mechanics, central element of the knowledge of every
physics student.

Nevertheless, the original Bohr statement about the term
complementarity (in  particular complementarity between wave and
particle behaviours) "to denote the relation of mutual exclusion
characteristic of the quantum theory with regard to an application
of the various classical concepts and ideas" \cite{Bohr} has been
recently subject of a wide debate and a paradigm where this
"mutual exclusion" must be interpreted in a weaker sense is
emerging. Incidentally, it must be noticed that the mutual
exclusion was thought by Bohr as a necessary condition for
internal consistency of his interpretation attributing a classical
behaviour to detection apparatuses \cite{Bohr,P1,au} (we do not
enter in this letter in the extensively debated and still
unclarified issue of measurement in quantum Mechanics, see Ref.s
\cite{au,PL} and Ref.s therein).

In particular a series of experiments with single photons and
atoms in interferometers  have shown how a gradual transition
takes place between the two aspects (wave and particle) \cite{au},
described by the Greenberger-Gisen inequality $P^2 + V^2 \le 1$,
where $P$ denotes the predictability of path (particle behaviour)
and $V$ the visibility of interference (wave behaviour). The
knowledge of "welcher weg " (which path) is therefore alternative
to coherence (and thus to the possibility of having interference)
with a smooth transition between a perfect "welcher weg" knowledge
and a $100 \%$ interference visibility. The extension of
complementarity to classical concepts of waves and particle in
every situation (including, also, for example tunnel effect or
birefringence) is nevertheless not contained in the mathematical
formalism of Quantum Mechanics and can be questioned \cite{PL}.

Furthermore, in interference experiments \cite{int,au} the use of
beam splitters (or similar devices) can be somehow modellized with
classical particles transmission and reflection \cite{P1}. In this
sense a large interest arouses an experiment \cite{jap} based on
the theoretical proposal of Ref. \cite{P0} where the coincidences
between photodetectors after a tunnel effect in a double prism of
single photons produced in Parametric Down Conversion are studied
(quantum version of the 1897 J.C. Bose's experiment \cite{PL}).
More in detail, a single photon arriving on two prisms separated
by a small distance (less than the photon wave length) can either
be totally reflected or tunnels through the gap. In the first case
it will be sent to a first detector, in the second case to another
one; coincidences (anticoincidences) between these two detectors
are then measured. The result of this experiment was the
observation of anticoincidences between detectors showing that
single photons both had performed a tunnel (wave behaviour) and
had been detected in only one of the detectors ("whelcher weg"
knowledge). Should the photon behave like a classical wave, part
of it would be reflected and a part would tunnel, originating
coincidences. If it would behave like a particle it would not be
able to tunnel. On the other hand, in Quantum Optics, the single
photon either tunnels or is reflected, thus it reveals itself as a
wave when it tunnels through the gap, but keeping its particle
behaviour indivisible and following a specific path. Experimental
results, according to the authors, lead, therefore, to an
agreement "with quantum optics, namely, light showed both
classical wave-like and particle-like pictures simultaneously" "in
contrast with conventional interpretation of the duality
principle".

In Ref. \cite{P1} this result was interpreted as an indication in
favour of de Broglie Bohm theory \footnote{For a recent experiment
aimed at testing this theory against standard quantum mechanics
see Ref. \cite{nos}.}, however this interpretation is debated
\cite{au}.

Nevertheless, the results of this experiment were questioned
\cite{Un,PL} as a case of "an insufficient statistical precision".
In particular in Ref. \cite{Un} it was noticed that the parameter
$\alpha = {N_c N \over N_1 N_2}$ (where $N_c$ denotes coincidence
counts, $N$ the number of gates where photons are counted and
$N_1$ and $N_2$ single detector counts), which should be $\ge 1$
for a classical source and
 $<< 1$ for PDC quantum states \cite{int} (strictly zero
in absence of background), was $\alpha \simeq 1.5 \pm 0.6$ for the
data of Ref. \cite{jap}. According to Ref. \cite{int,Un,PL} this
parameter $\alpha$ is the best discriminator between classical and
quantum states. The experimental precision of Ref.\cite{jap} was
therefore largely insufficient to discriminate between classical
and quantum light.

Considered the large relevance of these studies for the very
foundations of Quantum Mechanics, we have decided to realize a new
version of this experiment where the previous limitations are
overcome. In particular, we have realized an experiment where the
wave behaviour is related to birefringence, as suggested in Ref.
\cite{PL} by one of the authors of the original proposal
\cite{P0}.

Our scheme consists of a heralded single photon source based on
type I parametric fluorescence generated by an UV pump laser into
a non-linear crystal. This source is obtained by using the PDC
property that photons are produced in pairs emitted within few
femtoseconds conserving energy and momentum, i.e. $\nu_0 = \nu_1 +
\nu_2$ and $\vec{k}_0 = \vec{k}_1 + \vec{k}_2$ where indexes
$0,1,2$ denote the pump and the two PDC photons respectively.
Thus, the observation of a photon, after spatial and spectral
selection, in a first detector (D3) implies the presence of a
second photon on the conjugated direction with a fixed frequency.
The detection of the first photon is therefore used to open a
coincidence window where the second photon is expected to be
detected. Before detection this second photon crosses a
birefringent crystal where its path is split according to its
polarization: birefringence (and in particular the fact that
refractive indices are both larger than unity) is a typical
phenomenon explained only in terms of wave like propagation.
Finally, two single photon detectors (D1 and D2) are placed on the
two possible paths (for ordinary and extraordinary polarization).
The measurement of coincidences between these two last detectors
in the window opened after a count in the first detector  (D1)
allows \cite{PL}, in complete analogy to Mizobuchi and Ohtak\'e
experiment \cite{jap}, observation of corpuscolar properties of
the photon (specific path) together with wave ones
(birefringence). On the other hand, the use of a high intensity
source and of a simple scheme allow to overcome the low statistic
limitations of the previous experiment.

More in details, our set-up, see Fig.1, consists of a vertically
polarized Argon laser beam at 351 nm pumping a lithium iodate
crystal (5x5x5 mm) where type I PDC (i.e. horizontally polarized)
is produced. One photon of the PDC correlated pairs (at 633 nm) is
detected, after an iris and an interferential filter (4 nm FWHM)
by an avalanche single photon-detector (D3). The output of this
detector is used as trigger (start) of two Time to Amplitude
Converters (TAC). The observation of this photon guarantees,
thanks to the entanglement properties of PDC light \cite{mandel},
the presence of a single photon on the conjugated arm, namely
realizes a heralded single photon source. To the previous TACs are
then routed respectively (as stop) the signals obtained by the two
single photon detectors (D1 and D2) placed on the ordinary
($45^o$) and extraordinary ($135^o$) paths selected by a calcite
crystal placed on the conjugated direction (at 789 nm) to the
former one. Both detectors are preceded by an iris and an
interferential filter, 4 nm FWHM. As expected, clear coincidence
peaks (between photodections in detectors D3 and the ones in D1,
D2 respectively) were in fact observed sending TACs outputs to a
multichannel analyzer, testifying its operativity as heralded
single photon source.

Finally, the Single Channel Analyzer (SCA) outputs, giving the
number of photo-detection in detectors D1 and D2 respectively
arriving in a temporal window of 7 ns after a signal in detector
D3, are routed to an AND circuit giving the coincidence counts
($N_c$). No background subtraction is performed (background
contributes as a classical source). In this configuration a
logical AND between the valid starts of the two TACs (where the
start is the number of counts measured by D3) represents therefore
the number of gates (N). $N_c$ and $N$, together with the number
of counts in the 7ns temporal window measured by D1 ($N_1$) and D2
($N_2$), allow the evaluation of $\alpha = {N_c N \over N_1 N_2}$.

The results of our experiment for this parameter  in function of
the average single counts on trigger channel D3 (corresponding to
different attenuations of the pump laser beam), are shown in
Fig.2. The data are obtained with 500 acquisitions of 1 s per
point, except the one at 20000 counts/s obtained with 5000
acquisitions of 1 s. As expected, due to small background, data
are compatible with zero at low single counts values. At larger
values accidental random coincidences are not anymore negligible
and the measured $\alpha $ increases, remaining however largely
under unity in  the whole investigated region.
 The weighted average of the first three points $\alpha= 0.022 \pm
 0.019$ is (within almost one standard deviation) compatible with zero
 and differs from unity of more than 51 standard deviations.

This result is what is expected from quantum mechanics formalism,
 since type I PDC correlated pairs produced by a vertically
polarized pump are well described by a state $| H \rangle | H
\rangle $ ($H$ denoting horizontal polarization).  The state of
the second photon (idler) in the $45^o - 135^o$ polarization basis
is therefore given by the superposition $| \Psi_i \rangle = { | 45
\rangle + | 135 \rangle \over \sqrt 2} $ . The two components are
split on different paths by the birefringent calcite crystal $|
\Psi_i \rangle = { | 45, path 1 \rangle + | 135, path 2 \rangle
\over \sqrt 2} $. Thus, coincidences of detections on path 1 and
2, are strictly zero (a part of background).

In order to compare PDC single photon results with a classical
source, we have repeated the measurement by using an attenuated
He-Ne laser, emitting at 633 nm. In this case the gate (the start
of TACs) is given by a pulse generator with a trigger frequency
rate of 65 kHz (see Fig.3); the rest of the apparatus is the same
as before. In Fig.4 (similarly to Fig.2) we report the results
(with 500 acquisitions of 1 s per point) for this case as function
of the average single counts  of one of the detectors (i.e. of
laser power, attenuated by inserting neutral filters). Our
experimental average datum $\alpha = 0.9980 \pm 0.0022$ is in
perfect agreement (within one standard deviation) with the result
expected  for laser coherent light, i.e. \cite{int} $\alpha =1$.

For the sake of completeness we have also repeated the same
experiment with thermal light emitted by a tungsten lamp focalized
in the calcite crystal.

Following the formalism presented in Ref. \cite{mandel}, for
orthogonal polarizations ($P_1,P_2$), the correlation function $
\Gamma^{(2,2)}(r_1,r_2) = \langle I(r_1) I(r_2) \rangle$ (where
$r_1,r_2$ are the position of the detector 1 and 2 respectively
and $I$ is the intensity of the field) factorizes  $
\Gamma^{(2,2)}(r_1,r_2) = \langle I(r_1,P_1) \rangle \langle
I(r_2,P_2) \rangle$; therefore also in this case the expected
result is $\alpha =1$. Our experimental results (always with 500
acquisitions of 1 s per point), presented in Fig.5 in function of
average single counts of one of the detectors (i.e. of the lamp
intensity), give an average value $\alpha = 1.0010 \pm 0.0028$ in
perfect agreement with the former prediction.

Thus, in conclusion, the PDC single photon  result reported:
$\alpha= 0.022 \pm 0.019$ agrees with Quantum Optics expectations,
being largely below unity (more than 51 standard deviations) that
would be the result expected (and obtained) by using classical
sources.

In summary, we have realized a new version of the experiment
suggested in Ref. \cite{P0,PL}, which overcomes limitations
\cite{Un} of a previous similar experiment  \cite{jap} leading to
a conclusive answer to the questions raised in the original
theoretical paper.

Our results, in perfect agreement with the mathematical formalism
of quantum mechanics predictions, give a clear and conclusive
indication that wave particle duality must be considered in a weak
sense as a gradual disappearing of interference when "welcher weg"
indications are obtained and not in the original Bohr's form
asserting a complete exclusion of every wave and particle aspects.

 \vskip 0.5cm

{\bf Acknowledgments}

We thank P. Ghose for having pointed out to us the interest for
this experiment.

We acknowledge support of  MIUR (FIRB RBAU01L5AZ-002) and Regione
Piemonte.


\newpage

  \begin{figure}
   \begin{center}
   \begin{tabular}{c}
   \includegraphics[height=7cm]{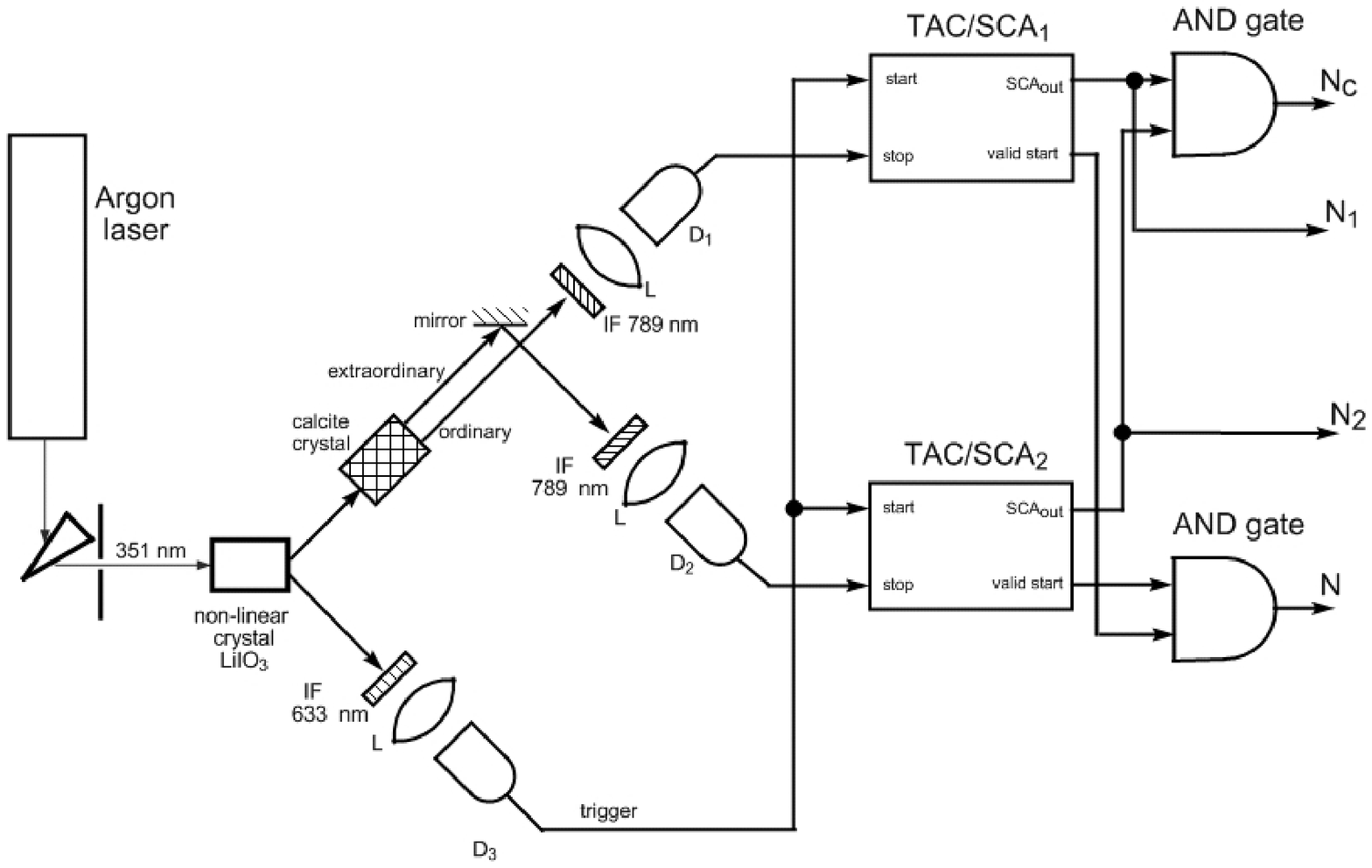}
   \end{tabular}
   \end{center}
   \caption[Fig. 1 ]
   { \label{fig::1} The experimental set-up. A vertically polarized Argon laser
beam at 351 nm pumps a lithium iodate crystal (5x5x5 mm) where
type I PDC (i.e. horizontally polarized) is produced. One photon
of the PDC correlated pairs (at 633 nm) is detected, after an
iris, a lens (L) and an interferential filter  (IF) by an
avalanche single photon-detector (D3) and used as start of two
Time to Amplitude Converters. To these TACs are  then routed (as
stop) the signals obtained by two single photon detectors (D1 and
D2) placed on the ordinary ($45^o$) and extraordinary ($135^o$)
paths selected by a calcite crystal placed on the conjugated
direction (789 nm) to the former one (both detectors preceded by
an iris, lens  and an interferential filter). The outputs of the
two TAC are then routed to an AND circuit giving coincidences
($N_c$). }
   \end{figure}

\newpage
\begin{figure}
   \begin{center}
   \begin{tabular}{c}
   \includegraphics[height=10cm]{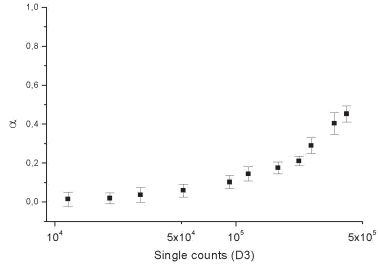}
   \end{tabular}
   \end{center}
   \caption[Fig. 1 ]
   { \label{fig::2} Values of the parameter $\alpha$ (see text) for heralded
single photons produced by PDC in function of the single counts of
trigger detector (intensity of the pump laser). }
   \end{figure}

\newpage
\begin{figure}
   \begin{center}
   \begin{tabular}{c}
   \includegraphics[height=7cm]{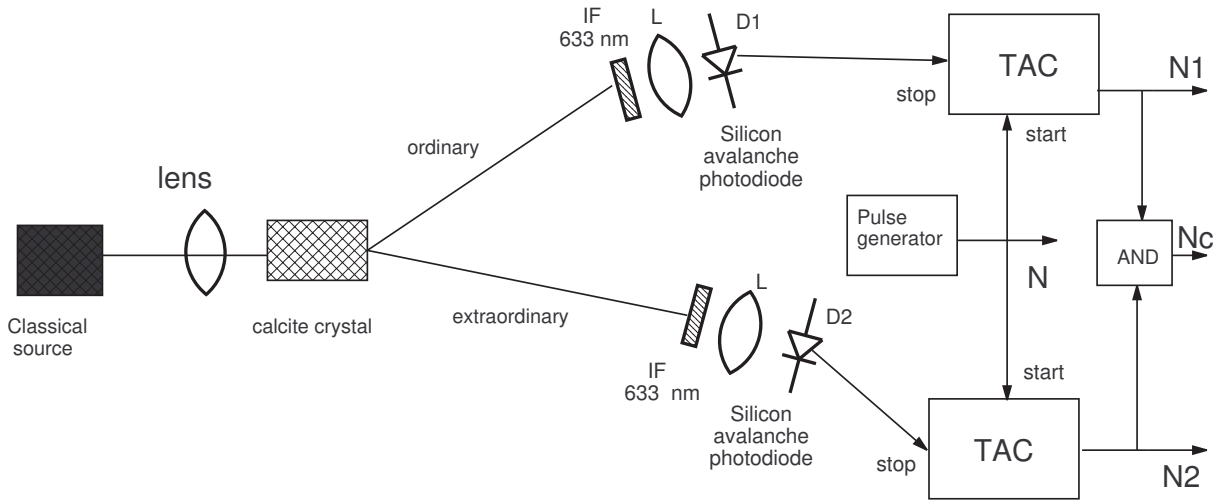}
   \end{tabular}
   \end{center}
   \caption[Fig. 1 ]
   { \label{fig::3} Experimental set-up for a classical source. The attenuated
classical source light is focalized into the calcite crystal
splitting ordinary and extraordinary rays. The two branches are
than measured by single photon detectors which are routed as stop
to two TACs. The start signal to TACs is given by a pulse
generator. The outputs of TACs feed an AND logical gate giving the
coincidence counts. }
   \end{figure}

\newpage
\begin{figure}
   \begin{center}
   \begin{tabular}{c}
   \includegraphics[height=10cm]{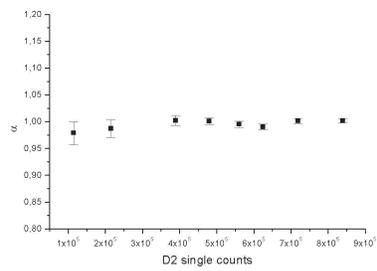}
   \end{tabular}
   \end{center}
   \caption[Fig. 1 ]
   { \label{fig::4} Values of the parameter $\alpha$ in function of the single
counts of one of the detectors for an He-Ne laser beam.}
   \end{figure}

\newpage
\begin{figure}
   \begin{center}
   \begin{tabular}{c}
   \includegraphics[height=10cm]{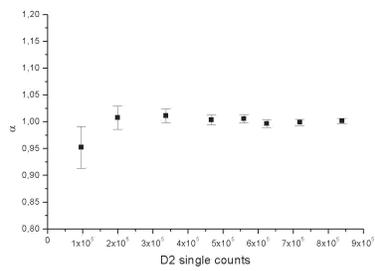}
   \end{tabular}
   \end{center}
   \caption[Fig. 1 ]
   { \label{fig::5} Values of the parameter $\alpha$ in function of the single
counts of one of the detectors for thermal light (tungsten lamp
emission) }
   \end{figure}

\end{document}